\documentclass[aps,preprint,preprintnumbers,amsmath,amssymb]{revtex4}
\usepackage{graphicx}
\usepackage{dcolumn}
\usepackage{bm}
\begin{document}

\title{Fabrication and Characterization of Modulation-Doped ZnSe/(Zn,Cd)Se (110) Quantum Wells: A New System for Spin Coherence Studies}
\author{K. C. Ku}
 %\altaffiliation[Also at ]{Department of Physics, The Pennsylvania State University.}%Lines break automatically or can be forced with \\ 
\author{S. H. Chun\footnote{Present address: Department of Physics, Sejong University, Seoul 1430747, Korea.}}
\author{W. H. Wang}
\author{W. Fadgen}
\author{D. A. Issadore}
\author{N. Samarth}\email{nsamarth@psu.edu}%
 
\affiliation{
Department of Physics, The Pennsylvania State University, University Park, PA 16802}

\author{R. J. Epstein}
\author{ D. D. Awschalom}
\affiliation{
Department of Physics, University of California, Santa Babara, CA 93106}.

\date{\today}

\begin{abstract}
We describe the growth of modulation-doped ZnSe/(Zn,Cd)Se quantum wells on (110) GaAs substrates. Unlike the well-known protocol for the epitaxy of ZnSe-based quantum structures on (001) GaAs, we find that the fabrication of quantum well structures on (110) GaAs requires significantly different growth conditions and sample architecture. We use magnetotransport measurements to confirm the formation of a two-dimensional electron gas in these samples, and then measure transverse electron spin relaxation times using time-resolved Faraday rotation. In contrast to expectations based upon known spin relaxation mechanisms, we find surprisingly little difference between the spin lifetimes in these (110)-oriented samples in comparison with (100)-oriented control samples.
\\
{\bf KEY WORDS} spin coherence, time-resolved Faraday rotation, ZnSe 
\end{abstract}

\maketitle

The idea of utilizing electron spin states in semiconductors for quantum information processing has attracted substantial attention in recent years \cite{awschalom,divincenzo95}. It is important in this context to develop a basic understanding of electron spin coherence in different semiconductors. Although direct measurements of the homogeneous spin coherence time ($T_2$) are difficult, time-resolved Faraday rotation (TRFR) measurements provide a measure of the inhomogeneous transverse spin lifetime (${T_2}^*$) which sets a lower limit on $T_2$. The most extensive measurements of TRFR have focused on n-GaAs, demonstrating that ${T_2}^*$ can be as long as 130 ns at low temperatures in bulk samples doped near the metal-insulator transition \cite{kikkawa98}. Although ${T_2}^*$ is shorter in modulation-doped quantum wells (QWs), it can be varied by tailoring the crystalline orientation of the QW: for instance, TRFR measurements of ${T_2}^*$ in (110)-oriented GaAs QWs yield significantly longer spin lifetimes compared with (100)-oriented samples \cite{salis01}.   These observations are qualitatively consistent with theoretical models \cite{lau01,lau04} for spin relaxation due to the D'yakonov-Perel (DP) mechanism \cite{dyakonov71, dyakonov86}. However, there is a need for additional experimental measurements in different materials systems to test the generality of these concepts. 

Here, we describe the development of modulation-doped (110) ZnSe/(Zn,Cd)Se QWs, a new heterostructure that can provide further insights into spin coherence in semiconductors. Our choice of material is motivated by earlier measurements showing relatively long spin coherence times in (001)-oriented ZnSe/(Zn,Cd)Se QWs and heterostructures \cite{kikkawa97, malajovich00a, malajovich00b, malajovich01}. Since  (110)-oriented ZnSe heterostructures are rarely documented in the literature (unlike their GaAs-based counterparts), we provide a detailed discussion of the growth conditions required for the molecular beam epitaxy (MBE) of ZnSe(110) QWs. We then demonstrate the fabrication of two-dimensional electron gases (2DEGs) of reasonable mobility in modulation-doped samples. Finally, we measure ${T_2}^*$ in these samples over a temperature range $4.2 K \leq T \leq 200$ K. To our surprise, we find that the spin lifetimes in these (110)-oriented samples do not differ significantly in comparison with measurements in (001)-oriented samples of comparable mobility. 

All the samples in this study are grown in an Applied EPI 620/930 dual-chamber MBE system with dedicated III-V and II-VI growth chambers. We use standard effusion cells containing elemental solid sources of Zn(6N), Se(6N), Cd(6N) As(7N), and Ga(7N) and a compound source of ZnCl$_2$ for n-doping of ZnSe. The growth is monitored \emph{in situ} by reflection high energy electron diffraction (RHEED) and the growth temperature is monitored by a thermocouple situated behind the substrate mounting block. During GaAs growth, the growth temperature is calibrated using a IRCON pyrometer. Substrates are mounted on molybdenum blocks using indium bonding. Samples are grown on both non-vicinal and vicinal (6$^\circ$ toward (111)B) (110) GaAs substrates. Following the growth, the samples are studied using atomic force microscopy (AFM), photoluminescence (PL) spectroscopy, Hall effect measurement and time-resolved Faraday rotation (TRFR).  The AFM measurements utilize a Nanoscope III AFM from Digital Instruments in tapping mode. The PL measurements are performed using 325 nm excitation from a He-Cd laser. The Hall measurements are carried out on mesa-etched Hall bars using dc techniques in an external magnetic field up to 8 T at temperatures between 340 mK and 4.2K. Electrical contacts are made using gold wire leads and indium solder in a forming gas (95\%N$_2$+5\%H$_2$) at 300$^\circ$C for 15 minutes. The TRFR measurements are performed in the Voigt geometry using optical pump-probe time-delay technique described elsewhere \cite{epstein02}. 

The standard approach to the epitaxial growth of ZnSe on (001) GaAs is well-documented: smooth epilayers are routinely grown at a substrate temperature $T_S \sim 250^\circ \rm{C} - 300^\circ \rm{C}$,  either directly on thermally-desorbed substrates or on a standard buffer layer of GaAs typically deposited in a dedicated III-V MBE chamber. Both vicinal and non-vicinal (001) GaAs substrates yield epitaxial ZnSe of comparable crystalline quality. However, we find that none of these standard approaches work for the epitaxy of ZnSe on (110) GaAs, leading instead to rough surfaces. For instance, Fig. 1(a) shows the result of growing a 700 nm thick layer ZnSe on a vicinal (110) GaAs substrate, after the deposition of a standard (110) GaAs buffer layer ($T_S \sim 400^\circ$C, As/Ga beam equivalent pressure ratio (BEPR) $\sim 100$\cite{wang83, neave93}). Under Se-rich conditions (Se/Zn BEPR=2), the RHEED shows 2D growth during epitaxy. However, AFM measurements indicate substantial morphological undulation elongated parallel to [$\bar{1}$10]. This surface structure is a one-dimensional equivalent of ``cross-hatch" and is created by enhanced growth over strain-relaxed regions due to lateral mass transport via surface diffusion\cite{andrews00}. 

We find that the growth of smooth ZnSe epilayers on (110) GaAs requires the following protocol: a thin (30 nm) epilayer of GaAs is first deposited on vicinal (110) GaAs, followed by another thin (150 nm) low-temperature GaAs epilayer grown at $T_S \sim 200^\circ$C with As/Ga BEPR = 15. This allows the subsequent deposition of ZnSe up to thicknesses of  $\sim 1 \mu$m with an order-of-magnitude decrease in surface roughness, as shown in Figure 1(b). We suggest that the low-temperature GaAs buffer improves the growth because it has an expanded lattice constant\cite{kaminska89}, hence reducing the lattice mismatch between ZnSe and GaAs. Table 1 summarizes the (110) optimal growth condition used in this study in comparison with (001) growth. Note that the II-VI growth temperatures quoted here are from the thermocouple reading.

We have used the growth protocol described above to fabricate modulation-doped QWs of ZnSe/Zn$_{1-x}$Cd$_x$Se with Cd composition up to $x = 0.3$. Each sample contains a strained 10 nm Zn$_{1-x}$Cd$_x$Se ($0.1 \leq x \leq 0.3$) QW and is symmetrically doped with 10 nm ZnSe spacer layers separating the QW from the 20 nm Cl-doped ZnSe doping layers. A 0.8 $\mu$m ZnSe buffer layer separates the QW structure from the GaAS substrate. At $T = 5$K, we observe PL linewidths in the range 10-20 meV, indicating QWs of reasonable quality (see Figure 2 inset).  Figure 2 shows Hall data taken at $T = 4.2$ K and $T = 340$ mK, respectively, in a perpendicular magnetic field $0 \leq B \leq 8$T. Clear quantum Hall plateaus are observed at both temperatures for $B > 3$T, indicating the presence of a 2DEG; however, closer analysis of the Hall effect data shows the existence of parallel conduction, preventing an accurate determination of the carrier density and mobility. The onset of quantum oscillations at $B \sim 3$T does however provide a rough estimate of the mobility $\mu \sim 3000 \rm{cm}^2$/V.s. The 2DEG carrier density can also be roughly estimated by assigning self-consistent filling factors to the center of the two plateau-like regions. For instance, for the data shown in Fig. 2, we estimate $n \approx  3 \times 10^{11} \rm{cm}^{-2}$ by assigning the filling factor $\nu = 3$ at $B = 4$T. At present, we do not understand the origin of the parallel conduction in these (110)-oriented samples since we can reproducibly fabricate 2DEG samples in (100)-oriented samples without a parallel conduction channel.    

We use TRFR to measure ${T_2}^*$ in (110) ZnSe/(Zn,Cd)Se 2DEGs, as well as in (001)-oriented 2DEGs of similar carrier density and mobility.  A train of 100 fs pulses from a mode-lock Ti:sapphire laser is split into pump and probe beams whose temporal delay, $\Delta t$, is adjusted using a mechanical delay line. The circularly polarized pump is tuned to the heavy-hole QW adsorption and generates spin-polarized electrons in the QW. We measure the Faraday rotation of the probe's linear polarization, which is proportional to the electron spin magnetization along the probe beam's direction \cite{crook97}. Sweeping $\Delta t$ while recording the Faraday rotation of the linearly polarized probe reveals the dynamics of electron spins, and provides both Larmor precession and transverse spin relaxation. We observe TRFR oscillations using an excitation wavelength of 475 nm, which is just above the absorption edge of the QWs. Figure 3(a) shows TRFR measurements at $T = 175$K obtained for a (110)-oriented sample with a carrier density of $n \approx 1 \times 10^{11} \rm{cm}^{-2}$ at $B = 0$~ and $B = 1$T. 

We estimate ${T_2}^* = 1.5$ns by fitting the exponential decay of the TRFR envelope amplitude; we also deduce the g-factor $g = 1.1$ from the periodicity of the oscillations. Note that the magnetic field dependence of ${T_2}^*$~ is very weak, suggesting little inhomogeneous dephasing of spins as they precess. This weak field dependence is observed in all 3 samples we have studied so far. We also observe a decrease of ${T_2}^*$~ with n-doping, as in earlier studies of both bulk and (001) QW samples \cite{kikkawa97}. Figure 3(b) depicts the temperature dependence of ${T_2}^*$~ for (110)-grown and (001)-grown ZnSe/(Zn,Cd)Se 2DEGs. Compared to the striking differences observed between the (001)- and (110)-orientations in GaAs QWs,  the data for these II-VI QWs do not show significant differences between the two crystalline orientations. We note that expectations based upon the DP spin relaxation mechanism would predict a $\sim 33 \%$ longer spin lifetime in the (110)-oriented samples compared to (100)-oriented samples (assuming a {\it comparable} mobility and carrier density)\cite{lau04}. Since the DP mechanism is expected to dominate other mechanisms at high temperatures, the high temperature data in Fig. 3 would appear to be at odds with these expectations. However, we caution that more systematic experimental studies will be necessary before drawing substantial conclusions.  

To summarize, we have described the growth conditions necessary for the MBE growth of (110)-oriented ZnSe/(Zn,Cd)Se 2DEGs. We find that it is essential to use a template consisting of a vicinal (110) GaAs substrate and a low-temperature GaAs buffer to enable the subsequent deposition of micron-thick epilayers of ZnSe with good crystalline quality. TRFR measurements of the transverse spin lifetime  in these (110)-oriented 2DEGs yield similar behavior to that measured in (001)-oriented 2DEGs of comparable mobility. These results differ significantly with similar measurements in GaAs-based 2DEGs where spin lifetimes in (110)-oriented samples are typically much longer compared with (001)-oriented samples. Our observations indicate a need for significant efforts to understand the spin relaxation mechanism in these II-VI materials. This work was supported by the DARPA QUIST program.

\newpage
\begin{center}{\bf Table Captions}\end{center}
\begin{table}[h]
\begin{center}
\caption{Comparison of optimal MBE growth conditions for ZnSe(110) and ZnSe(001) on GaAs. For (110) growth, vicinal GaAs substrates are used. The growth temperatures for ZnSe are obtained by thermocouple readings.}
\label{sum}
\end{center}
\end{table}

\newpage
\begin{center}{\bf Figure Captions}\end{center}
\begin{figure}[h]
\begin{center}
\caption{Surface profiles obtained by AFM of (a) 700 nm ZnSe epilayer on a standard GaAs layer (RMS roughness = 46 nm) and (b) 1 $\mu$m ZnSe epilayer on a LTGaAs layer (RMS roughness=1 nm). For clarity, the scan in (b) is shown over a smaller region than (a);  we note that the order of magnitude decrease in RMS roughness is found over scan areas comparable in size to (a). Both samples are grown on vicinal substrates at optimal (110) growth temperature.}
\label{afm}
\end{center}
\end{figure}

\begin{figure}[!h]
\begin{center}
\caption{Hall effect data at 4.2 K (dashed line) and 340 mK (solid line) in the (110)-oriented 2DEG whose structure is described in the text (PL peak = 2.7 eV at T = 5 K, FWHM = 17 meV, Cl cell temperature=180$^\circ$C). Based on the positions of the quantum Hall plateaus, we estimate a sheet density $n \sim 3 \times 10^{11} \rm{cm}^{-2}$ and mobility $\mu \sim 2000$cm$^2$/V.s at $T = 4.2$K. The inset shows the PL spectrum from another (110)-oriented 2DEG at $T = 5$ K (Cl cell temperature = 165$^\circ$C).}
\label{hall}
\end{center}
\end{figure}

\begin{figure}[!h]
\begin{center}
\caption{(a) TRFR in a (110)-oriented ZnSe/(Zn,Cd)Se 2DEG with n$\approx$3$\times$10$^{11}$cm$^{-2}$ at $B = 0$ T and $B = 1$ T. Data taken at $T = 175$ K with pump power of $\sim 1$ mW and a pump energy of 2.72 eV. (b) Temperature dependence of $T^*_2$ of a ZnSe(110) 2DEG with $n \approx 3 \times 10^{11} \rm{cm}^{-2}$ and a ZnSe(001) 2DEG with $n \approx 5 \times 10^{11} \rm{cm}^{-2}$.}
\label{trfr}
\end{center}
\end{figure}

\newpage
\begin{center}
\begin{tabular}{c|c|c}
 & \scriptsize{Growth Temperature}&\scriptsize{BEPR}\\ \hline
\scriptsize{(110)}&  \scriptsize{GaAs=400$^\circ$C, LTGaAs=200$^\circ$C}&\scriptsize{As/Ga=15 (100 for LTGaAs)}\\ 
&\scriptsize{ZnSe=325$^\circ$C}&\scriptsize{Se/Zn=2}\\ \hline
\scriptsize{(001)}& \scriptsize{GaAs=580$^\circ$C}&\scriptsize{As/Ga=15}\\
&\scriptsize{ZnSe=375$^\circ$C}&\scriptsize{Se/Zn=2}
\label{summary}
\end{tabular}
\end{center}
\vspace{2in}
Ku \emph{et al}-Table 1
\newpage

\begin{center}
\includegraphics[scale=0.6]{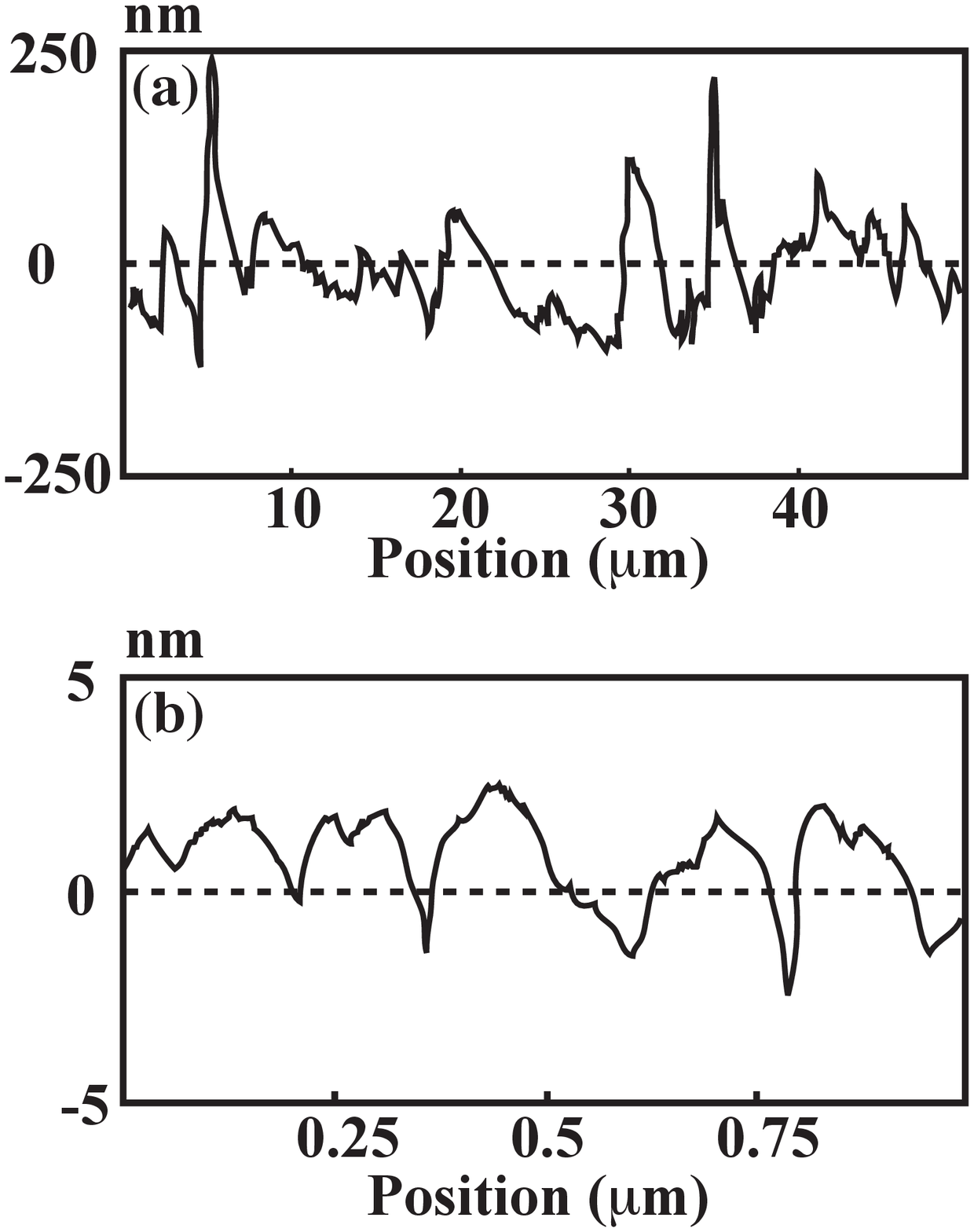}
\end{center}
Ku \emph{et al}-Figure 1
\newpage

\begin{center}
\includegraphics[scale=0.6]{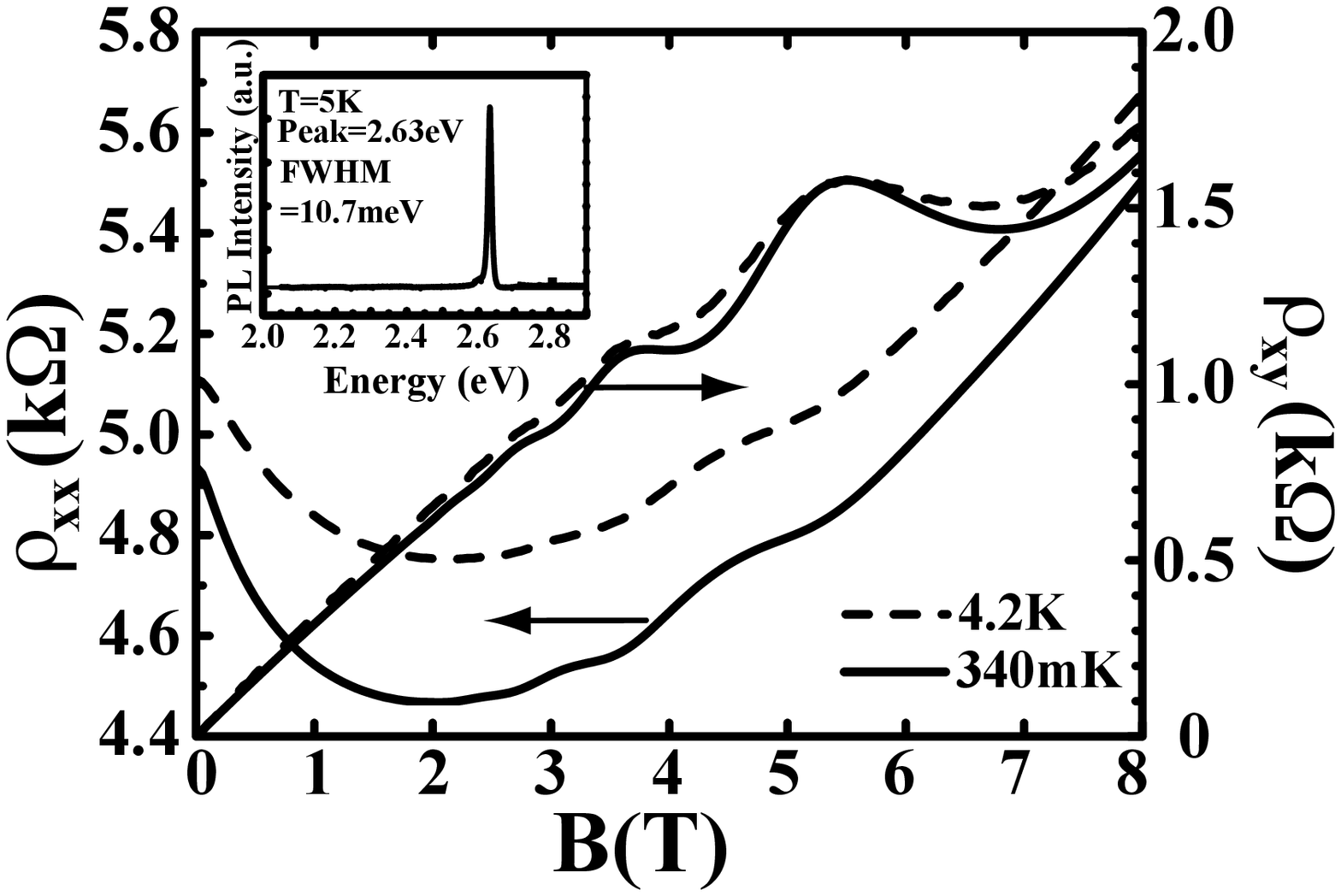}
\end{center}
Ku \emph{et al}-Figure 2
\newpage

\begin{center}
\includegraphics[scale=0.6]{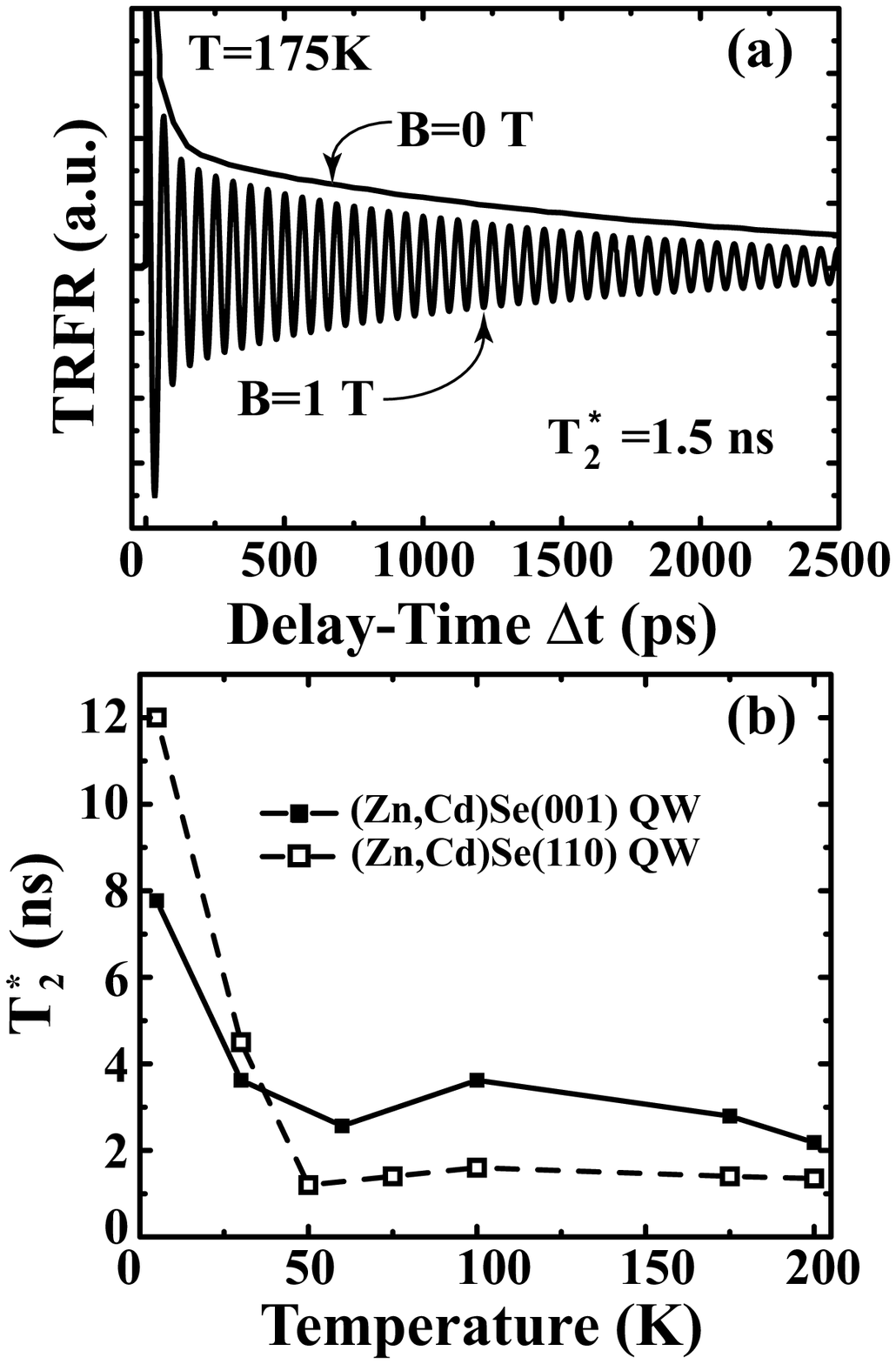}
\end{center}
Ku \emph{et al}-Figure 3

\end{document}